\documentclass[9pt,twocolumn,twoside]{osajnl}

\journal{ol} 

\setboolean{shortarticle}{true}

\title{Precision determination of the ground-state hyperfine splitting of trapped ${}^{\textbf{113}}$Cd${}^{\textbf{+}}$ ions}

\author[1]{S. N. Miao}
\author[1,*]{J. W. Zhang}
\author[2]{H. R. Qin}
\author[1]{N. C. Xin}
\author[1]{J. Z. Han}
\author[1,2,*]{L. J. Wang}

\affil[1]{State Key Laboratory of Precision Measurement Technology and Instruments, Key Laboratory of Photon Measurement and Control Technology of Ministry of Education, Department of Precision Instrument, Tsinghua University, Beijing 100084, China}
\affil[2]{Department of Physics, Tsinghua University, Beijing 100084, China}

\affil[*]{Corresponding authors: zhangjw@tsinghua.edu.cn, lwan@tsinghua.edu.cn}

\begin{abstract}
We measured the ground-state hyperfine splitting of trapped ${}^{\textbf{113}}$Cd${}^{\textbf{+}}$ ions to be 15199862855.02799(27) Hz with a fractional uncertainty of $\pmb{1.8\times10^{-14}}$. The ions were trapped and laser-cooled in a linear quadrupole Paul trap. The fractional frequency stability was measured to be $\pmb{{4.2} \times 10^{-13}/\sqrt{\tau}} $, obtained from Ramsey fringes of high signal-to-noise ratio and taken over a measurement time of nearly 5 hours, which is close to the short-term stability limit estimated from the Dick effect. Our result is consistent with previous reported values, but the measurement precision is four times better than the best result obtained to date.
\end{abstract}

\setboolean{displaycopyright}{true}

\begin{document}

\maketitle

Microwave atomic clocks have been playing an important role across many areas such as time standards, telecommunications, and deep space exploration \cite{burt2021demonstration}. Those based on trapped ions have featured alternate schemes that have exploited the compactness, high transportability, and zero-gravity operating capabilities of the clock. To date, ${}^{113}\mathrm{{Cd}}^{+}$, ${}^{171}\mathrm{{Yb}}^{+}$ and ${}^{199}\mathrm{{Hg}}^{+}$ ions have been employed in the development of microwave atomic clocks \cite{miao2015high,phoonthong2014determination,mulholland2019laser,berkeland1998laser,burt2021demonstration}. For instance, the laser-cooled ${}^{199}\mathrm{{Hg}}^{+}$ ion clock at the National Institute of Standards and Technology (NIST) achieved a frequency instability of $3.3(2) \times 10^{-13}/\sqrt{\tau} $ and a fractional frequency uncertainty of $1 \times 10^{-14} $ \cite{berkeland1998laser}. In a report by Phoonthong and colleagues \cite{phoonthong2014determination}, the short-term frequency instability of their Yb-ion microwave clock reached $2.09 \times 10^{-12}/\sqrt{\tau} $ with a fractional frequency uncertainty of $3.16 \times 10^{-14} $. 

Cadmium microwave ion clocks offer promising performances because of their large ground-state hyperfine splitting frequency (15.2 GHz) and their special energy level structure. Only one laser at 214.5 nm suffices in realizing laser cooling, pumping, and detection. Therefore, ${}^{113}\mathrm{{Cd}}^{+}$ clocks have significant potential in miniaturization. However, the literature contains only a few reports concerning such clocks. In 2006, researchers from the Jet Propulsion Laboratory (JPL) measured the ground-state hyperfine splitting frequency of ${}^{113}\mathrm{{Cd}}^{+}$ to be 15199862855.0(2) with a fractional frequency uncertainty of $1.3 \times 10^{-11} $ \cite{jelenkovic2006high}. Since 2010, our team at Tsinghua University has been committed to developing a microwave ion clock based on laser-cooled ${}^{113}\mathrm{{Cd}}^{+}$ ions. In 2015, we measured the ground-state hyperfine interval, the result being consistent with that of the JPL researchers \cite{miao2015high}. The precision was improved to $6.6 \times 10^{-14} $, the measured short-team frequency instability being $6.1 \times 10^{-13}/\sqrt{\tau} $.

In this letter, we report a more stable and more accurate microwave ion clock based on laser-cooled ${}^{113}\mathrm{{Cd}}^{+}$ ions. The ground-state hyperfine splitting frequency of ${}^{113}\mathrm{{Cd}}^{+}$ was determined to be 15199862855.02799(27) Hz with a fractional frequency uncertainty of $1.8 \times 10^{-14} $, the fractional frequency stability being $4.2 \times 10^{-13}/\sqrt{\tau} $. To the best of our knowledge, this cadmium-ion microwave ion clock currently has the best precision and stability.

\begin{figure}[h!]
\centering
\includegraphics[width=8.5cm]{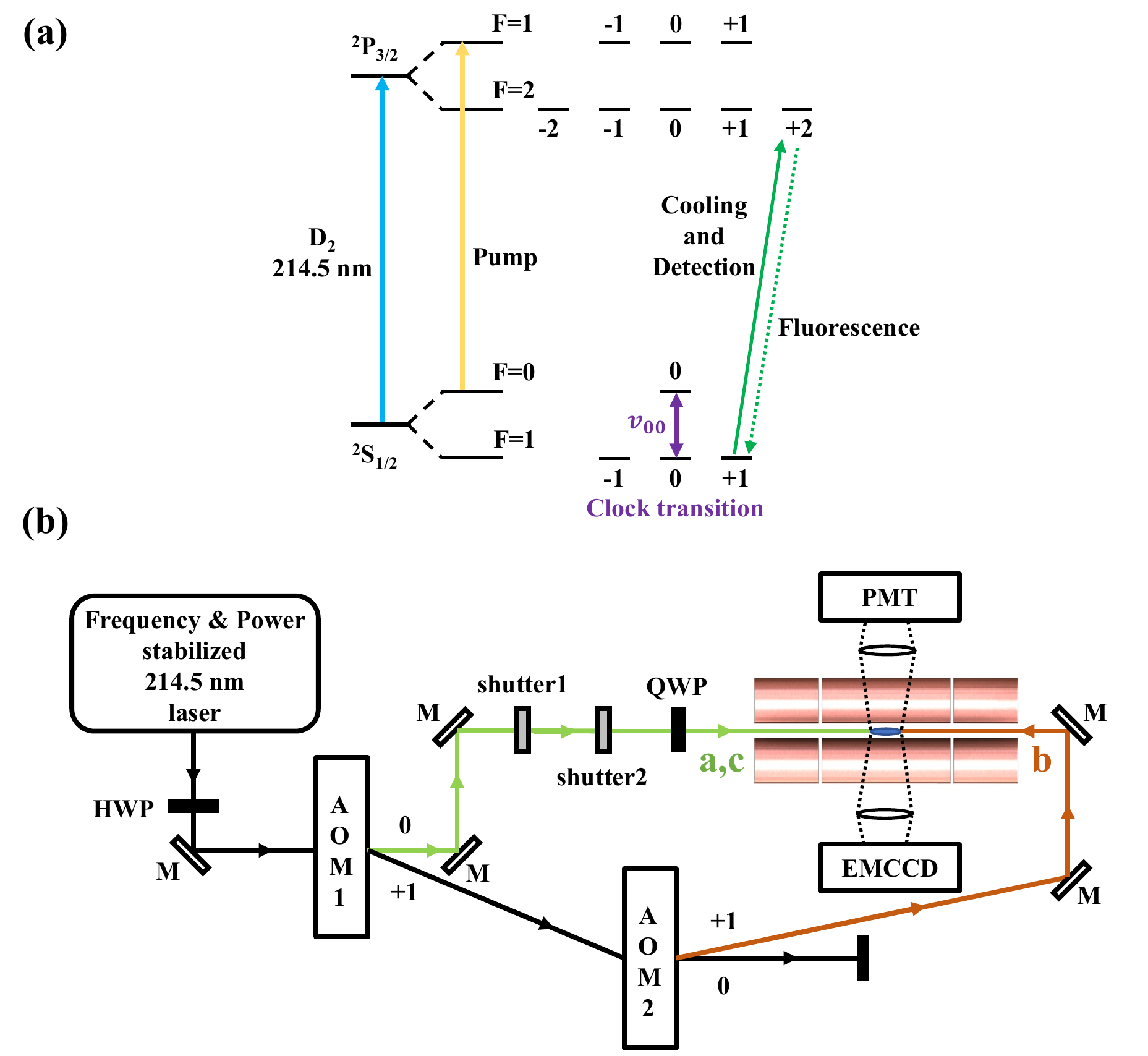}
\caption{\label{fig:fig_01} (a) Relevant energy level of ${}^{113}\mathrm{Cd}^{+}$(not to scale) and (b) schematic of the experiment system. M, mirror; HWP, half-wave plate; QWP, quarter-wave plate; AOM, acousto-optic modulator; PMT, photomultiplier tube detector; EMCCD, electron-multiplying charge coupled device. Laser beams a and c indicate cooling and detection,respectively, controlled by shutter2. Laser beam b refers to the pump beam. The cooling and detection laser beams are $\sigma$ polarization, whereas the pump laser is $\pi$ polarization.}
\end{figure}

The relevant energy level structure of ${}^{113}\mathrm{Cd}^{+}$ and our current system setup are shown in Fig. \ref{fig:fig_01}(a) and Fig. \ref{fig:fig_01}(b), respectively. The ions are cooled and detected by a 214.5-nm red detuned laser via the cycling transition between the states of $\left|^{2} S_{1 / 2}, F=1, m_{F}=1\right\rangle$ and $\left|^{2} P_{3 / 2}, F=2, m_{F}=2\right\rangle$. During laser cooling, ions stay at the energy level of $\left|^{2} S_{1 / 2}, F=0\right\rangle$ because of the relatively small frequency difference between the states of $\left|^{2} P_{3 / 2}, F=1\right\rangle$ and $\left|^{2} P_{3 / 2}, F=2\right\rangle$. Therefore, a cooling laser with circular polarization connecting two Zeeman sublevels of the cycling transition is applied. In addition, microwave radiation of 15.2 GHz is applied to repump the ions from the dark state of $\left|^{2} S_{1 / 2}, F=0\right\rangle$. Because the hyperfine splitting of $\left|^{2} P_{3 / 2}\right\rangle$ is only 800 MHz, the pump laser beam is generated by blue-shifting the cooling laser beam using acousto-optic-modulators (AOM). The frequency difference between the states of $\left|^{2} S_{1 / 2}, F=0, m_{F}=0\right\rangle$ and $\left|^{2} S_{1 / 2}, F=1, m_{F}=0\right\rangle$ is the chosen transition frequency, which is approximately 15.2 GHz.

The 214.5-nm laser system is a frequency-quadrupled, tunable diode laser system (TA-FHG Pro, Toptica, Inc). The output power of this laser is about 10 mW; both frequency and power are stabilized [Fig. \ref{fig:fig_01}(b)]. The cooling, pump, and detection laser beams are controlled by AOM-1, AOM-2, and shutter2. An attenuator plate with 10\% transmittance of the 214-nm laser beam is installed on shutter2 so that a beam power of order a hundred milliwatts is available for detection. The fluorescence signal from the ions is collected by a lens and detected by a photomultiplier tube (PMT) in photon-counting mode. Located on the opposite side of this tube, an electron-multiplying charge coupled device (EMCCD) images the structure of the ion cloud .

During the experiment, nearly ten thousand cadmium ions are trapped in a linear quadruple Paul trap. Details of the ion trap are given in \cite{zhang2014toward}; here, a brief description suffices. The linear ion trap consists of four parallel cylindrical electrode rods of radius 7.1 mm. The minimum distance from the nodal line of the trap to the electrode surface is 6.2 mm. The trap is driven by a radio frequency (rf) driver; the associated field has an amplitude of 250 V and frequency of 1.962 MHz. The DC voltage applied across the two endcaps is 10 V. The ion-trap construction is horizontally mounted in a stainless-steel vacuum chamber. The pressure in the vacuum chamber is maintained below $1.5\times10^{-11}$ mbar by an ion pump. A high-performance magnetic shield barrel composed of five layers of permalloy and a layer of soft-iron around the vacuum chamber is used to minimize stray magnetic fields. From finite-element simulations, the shielding coefficients at the center of the magnetic shield barrel in the south-north, east-west, and up-down directions were estimated to be 460000, 57000, and 970, respectively \cite{han2021toward}. In addition, the barrel is motor driven to move up and down. During the experiment, the barrel is raised so that the ion trap is positioned in the center of the barrel. After active shielding, the magnetic field fluctuation in the center of the ion trap may be reduced to less than 0.1 nT, which is much better than that in \cite{miao2015high}. To make the system more compact, three pairs of Helmholtz coils were fixed around the vacuum chamber to generate a static magnetic field to split the Zeeman sublevels and to compensate for any residual magnetic field. Moreover, to reduce electrical noise, more stable current sources with a short-term current stability of $4.6 \times 10^{-6}/\sqrt{\tau} $ were used to power the Helmholtz coils.

\begin{figure}[h!]
\centering
\includegraphics[width=8.5cm]{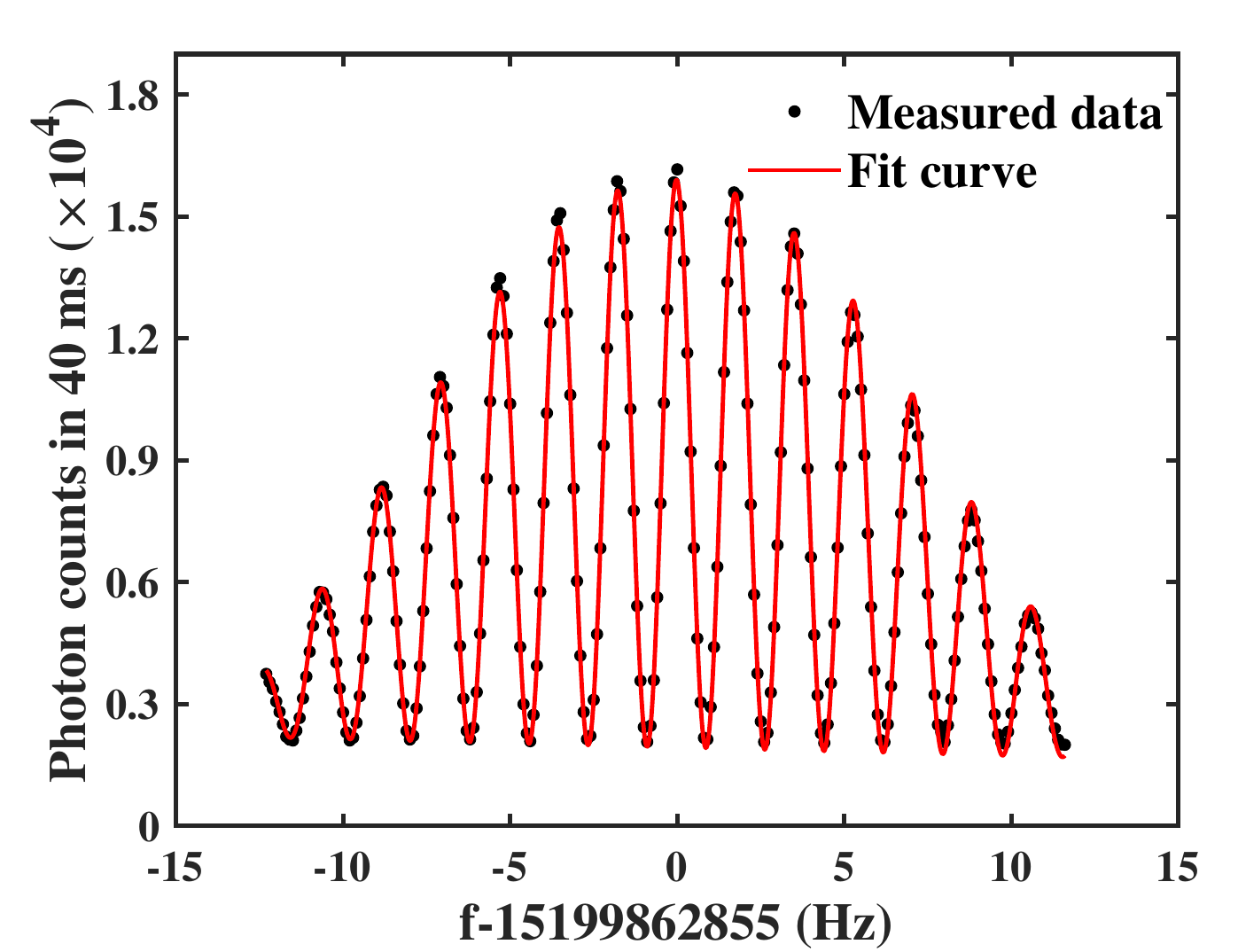}
\caption{\label{fig:fig_02} Typical high-SNR Ramsey fringe pattern obtained with a 500-ms cooling time, a 50-ms pump time, two phase-coherent microwave pulses of 60 ms duration separated by a free evolution time of 500 ms, and a fluorescence signal integration time of 400 ms. The scanning step is 0.1 Hz. The solid line is a nonlinear fit obtained from the experimental data.}
\end{figure}

Initially, using Ramsey's method of separated oscillation fields, we measured the spectra of the 0-0 ground-state hyperfine transition of ${}^{113}\mathrm{Cd}^{+}$. With a cooling time of 500 ms and a free evolution time of 500 ms, Ramsey fringe patterns were obtained(see Fig. \ref{fig:fig_02} for details). In the experiment, we scanned the microwave frequency $f$ over a $\pm 12 \mathrm{~Hz}$ range around the center frequency $f_0$. A nonlinear curve fitting of the measured data was performed using the expression for the magnetic dipole transition probability (Fig. \ref{fig:fig_02}, solid line). The high signal-to-noise ratio (SNR) of the transition signals indicates that the ions in the trap are very stable, which is essential in providing a high stability performance. 

Based on the Ramsey fringes with high-SNR, the closed-loop locking operation of the ${}^{113}\mathrm{Cd}^{+}$ microwave ion clock was carried out. The principle behind this closed-loop locking may be found in \cite{miao2015high}. The local oscillator (LO) is an oven controlled crystal oscillator (OCXO) with a frequency instability of $1.36\times10^{-13}/\mathrm{s}$. The LO produces two 10-MHz outputs, one of which is fed to a microwave synthesizer (8257D, Agilent) to generate the 15.2 GHz signal and the other is connected to the Phase Noise Analyzer (TSC 5120A, Timing Solutions Corp.) for comparison with the output from a hydrogen clock. 

\begin{figure}[t!]
\centering
\includegraphics[width=8.5cm]{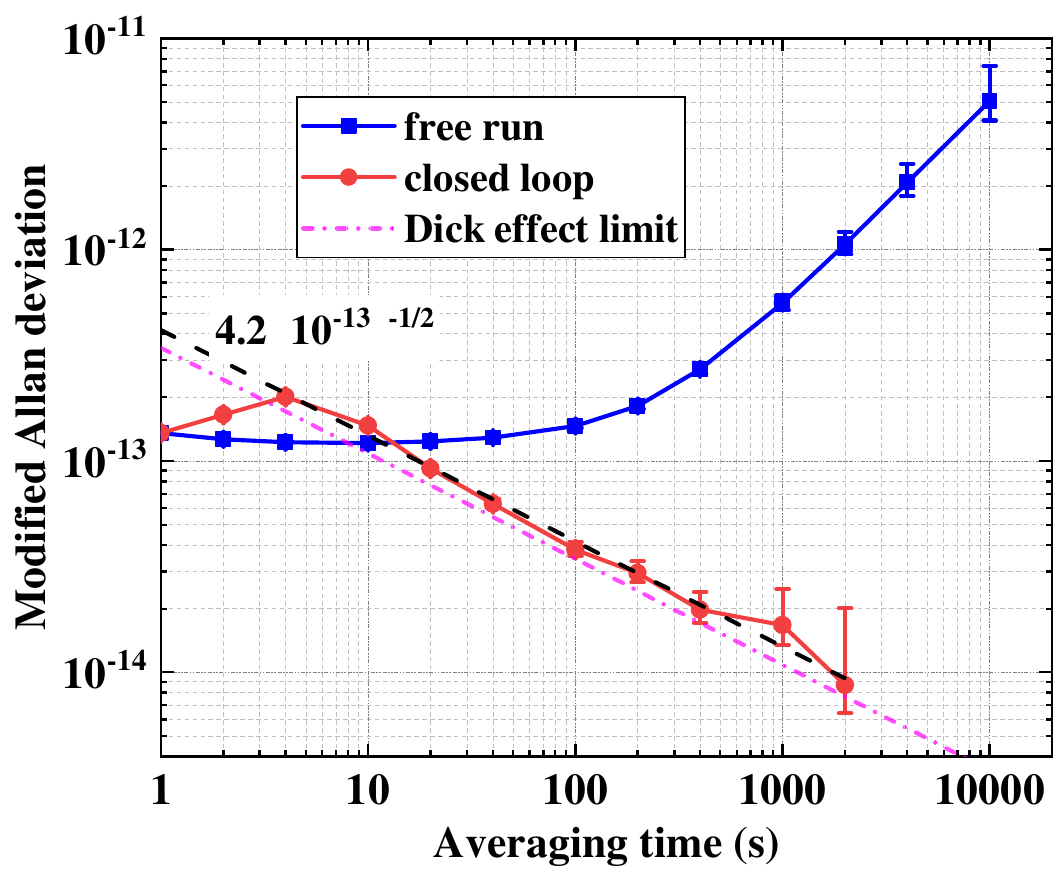}
\caption{\label{fig:fig_03} Modified Allan deviations of the ${}^{113}\mathrm{Cd}^{+}$ microwave ion clock. The solid blue line is a fitting of data from the free-running OCXO, the solid red line is that after closed-loop locking, and the pink dotted line signifies the Dick effect limit. The short-term fractional frequency stability of the clock is estimated to be $4.2 \times 10^{-13}/\sqrt{\tau} $.}

\end{figure}

From the measurement results (Fig. \ref{fig:fig_03}), the frequency stabilities for the free-running OCXO (solid blue line) and that after the closed-loop locking (solid red line) were determined. The loop time is 3.18 seconds, which corresponds to the time constant. Allan deviations longer than the time constant stem from the cadmium ions of the clock; those shorter than the time constant stem from the LO itself. After a long duration making comparisons, the fractional frequency stability of the ${}^{113}\mathrm{Cd}^{+}$ microwave ion clock was estimated to be $4.2 \times 10^{-13}/\sqrt{\tau} $, which reduces by nearly half that of the previous one \cite{miao2015high}. Under the current experimental parameters, the fractional frequency stability limit due to the Dick effect is $3.43 \times 10^{-13}/\sqrt{\tau} $, which implies that the noise of the LO is already the main factor limiting the short-term stability of the ${}^{113}\mathrm{Cd}^{+}$ microwave ion clock \cite{jian2015dick}.

The transition frequency $\nu_{00} (B_{0})$ between the states of $\left|^{2} S_{1 / 2}, F=0, m_{F}=0\right\rangle$ and $\left|^{2} S_{1 / 2}, F=1, m_{F}=0\right\rangle$ was also measured precisely to be 15199862856.63519(26) Hz. Considering that the frequency measurement is referenced against an active hydrogen clock, and that the fractional frequency difference between this active hydrogen clock and coordinated universal time (UTC) is $7.9(3)\times10^{-14}$, measured by the method of GPS common-view. The difference between UTC and the primary frequency standards (PFS) is $-0.03(14)\times10^{-15}$ \cite{BIPM}. Therefore, the transition frequency is corrected to be
\begin{equation}
\nu_{00} (B_{0})=15199862856.63399(26) \ \mathrm{Hz} 
\label{equ:equ_01}
\end{equation}
To obtain the 0-0 ground-state hyperfine transition frequency of ${}^{113}\mathrm{Cd}^{+}$, the systematic frequency shifts have to be evaluated carefully.

\begin{figure}[t!]
\centering
\includegraphics[width=8.5cm]{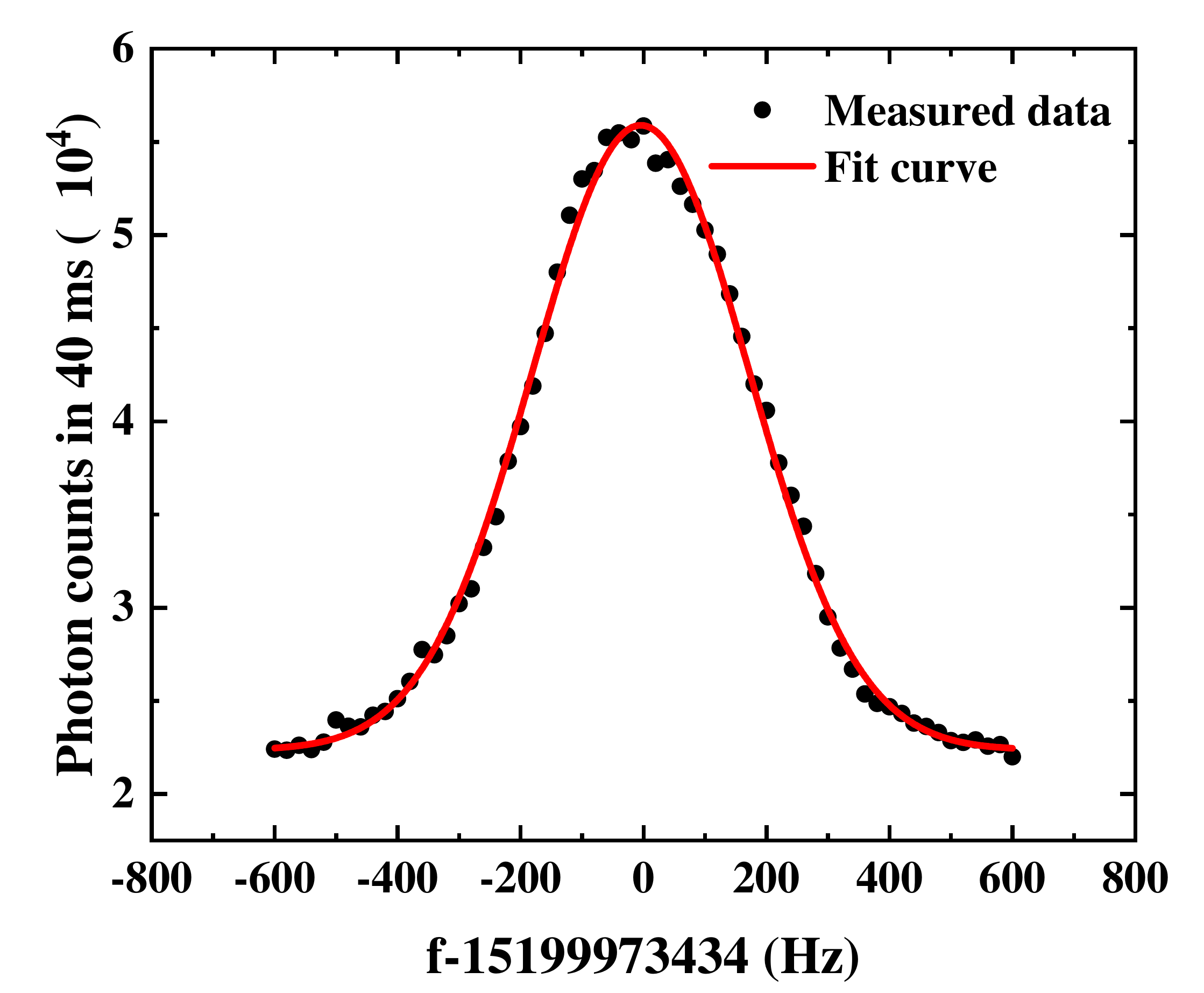}
\caption{\label{fig:fig_04} Rabi fringe of the magnetic-field-sensitive transition $\nu_{01} (B_{0})$ obtained using a  $\pi$ microwave pulse time of 40 ms. The power of the microwave pulse is -15 dBm; the scanning step length is 20 Hz.}
\end{figure}

According to the Breit-Rabi formula, the transition frequency $\nu_{00} (0)$ between states $\left|^{2} S_{1 / 2}, F=0, m_{F}=0\right\rangle$ and $\left|^{2} S_{1 / 2}, F=1, m_{F}=0\right\rangle$ at zero magnetic field is expressed as
\begin{equation}
\nu_{00}(0)=\sqrt{\nu_{00}\left(B_{0}\right)^{2}-\left(\frac{g_{J}-g_{I}}{g_{J}+g_{I}}\right)^{2}\left(\nu_{01}-\nu_{0-1}\right)^{2}}
\label{equ:equ_02}
\end{equation}
where $g_I$=$0.6223009(9)\times10^{-3}$ \cite{spence1972optical} and $g_J$=2.002291(4) \cite{yu2020ground} are the nuclear and electronic Land\'{e} g-factors, respectively, and $\nu_{01}$ and $\nu_{0-1}$ denote the respective frequencies of the magnetic-field-sensitive transitions $\left.\left|^{2} S_{1 / 2}, F=0, m_{F}=0\right\rangle\leftrightarrow|^{2} S_{1 / 2}, F=1, m_{F}=1\right\rangle$ and $\left.\left|^{2} S_{1 / 2}, F=0, m_{F}=0\right\rangle\leftrightarrow|^{2} S_{1 / 2}, F=1, m_{F}=-1\right\rangle$. For a typical Rabi fringe of one of the magnetic-field-sensitive transitions (Fig. \ref{fig:fig_04}), the microwave frequency is scanned over $\pm 600 \mathrm{~Hz}$ interval around the center frequency; note that the fringe has been narrowed by 35 times compared with that presented in \cite{miao2015high}. A measurement of the central frequencies of two Rabi fringes yielded a value for the frequency difference of $\left|\nu_{01}-\nu_{0-1}\right|$= 221158(3) Hz. According to Eq. (\ref{equ:equ_02}), the second-order Zeeman frequency shift (SOZS) is $105720\times10^{-15}$ with an uncertainty of $3\times10^{-15}$.

The second-order Doppler frequency shift (SODS), originating from the motion of the ions and estimating using the model for microwave atomic clocks \cite{prestage1999higher}, is given by
\begin{equation}
\frac{\delta \nu_{00}^{S O D S}}{\nu_{00}}=-\frac{3 k_{B} T}{2 M c^{2}}\left[1+\frac{2}{3}\left(N_{d}^{K}\right)\right],
\end{equation}
where $k_B$ denotes the Boltzmann constant, $T$ the temperature of ions, $M$ the mass of the ion, $c$ the vacuum speed of the light, and $N_{d}^{K}$ one of the key parameters related to the configuration of the trap and the state of the ion cloud. The temperature of ${}^{113}\mathrm{Cd}^{+}$ ions was measured to be 654(32) mK by scanning the Doppler broadening of the $D_2$ line. The value of $N_{d}^{K}$ was determined to be 122(3), obtained from molecular dynamics (MD) simulations; the details of our simulation model appear in \cite{xin2021research}. Thus, the fractional SODS is estimated to be $-65.9(3.6)\times10^{-15}$.

For blackbody radiation, the frequency shift includes contributions from the Stark (BBRS) and Zeeman (BBRZ) effects. For ${}^{113}\mathrm{Cd}^{+}$, assuming room temperature ($300 \pm 10$ K), the BBRS is estimated to be $-1.815(102)\times10^{-16}$ \cite{yu2017estimation}, and the BBRZ to be $-9.81(67)\times10^{-18}$ \cite{han2019theoretical}. Hence, it is safe to ignore the frequency uncertainty of the Blackbody radiation shift at this stage.

The altitude of the cadmium-ion microwave ion clock in our laboratory is measured to be $43\pm1$ m obtained using a commercial geodetic global positioning system (GPS) receiver and a laser rangefinder. Therefore, the fractional frequency shift resulting from the gravitational redshift is estimated to be $4.7(1)\times10^{-15}$.

The pressure frequency shift is negligible for an ultra-high vacuum below $1.5\times10^{-11}$ mbar, assuming the pressure shift coefficient is of the same order as for ${}^{171}\mathrm{Yb}^{+}$ \cite{park2007171yb+} and ${}^{199}\mathrm{Hg}^{+}$ \cite{chung2004buffer}. The frequency shift for light is also ignored because all laser beams were blocked by optical shutters during microwave interrogations.

\begin{table}[t!]
\centering
\caption{\bf Estimated Systematic Frequency Shifts and Uncertainties}
\begin{tabular}{ccc}
\hline
Shift & Magnitude of effect & Uncertainty \\
 &($\times10^{-5}$ Hz)&($\times10^{-5}$ Hz)\\
\hline
SOZS & 160693 & 5 \\
SODS & -100 & 5 \\
BBRS & $\textless$0.3 & $\textless$0.02 \\
BBRZ & $\textless$0.02 & $\textless$0.01 \\
Gravitational redshift & 7 & $\textless$0.2 \\
Pressure shift & 0 & $\textless$0.1 \\
Light shift & 0 & 0 \\
Total & 160600 & 7 \\
\hline
\end{tabular}
  \label{tab:tab_01}
\end{table}
\begin{table}[t!]
\centering
\caption{\bf Comparison of Different Measurement Results}
\begin{tabular}{clc}
\hline
Ref. & Measurement result (Hz) & Uncertainty \\
\hline
\cite{tanaka1996determination} & 15199862858(2) & $1.3 \times 10^{-10}$ \\
\cite{jelenkovic2006high} & 15199862855.0(2) & $1.3 \times 10^{-11}$ \\
\cite{zhang2012high} & 15199862854.96(12) & $7.9 \times 10^{-12}$ \\
\cite{wang2013high} & 15199862855.0125(87) & $5.7 \times 10^{-13}$ \\
\cite{miao2015high} & 15199862855.0287(10) & $6.6 \times 10^{-14}$ \\
This Letter & 15199862855.02799(27) & $1.8 \times 10^{-14}$ \\
\hline
\end{tabular}
  \label{tab:tab_02}
\end{table}

All of the systematic frequency shifts and corresponding frequency uncertainties discussed above are listed in Table \ref{tab:tab_01}. Finally, the 0-0 ground-state hyperfine splitting frequency of ${}^{113}\mathrm{Cd}^{+}$ is determined to be
\begin{equation}
\nu_{00} (0)=15199862855.02799(27) \ \mathrm{Hz}
\label{equ:equ_04}
\end{equation}

Table \ref{tab:tab_02} lists all the measurement results of the 0-0 ground-state hyperfine splitting frequency for ${}^{113}\mathrm{Cd}^{+}$ found in the literature. Our new measurement result is consistent with previous results. To the best of our knowledge, the measurement accuracy is currently the highest among all ${}^{113}\mathrm{Cd}^{+}$ microwave atomic clocks.

We have described a highly stable and accurate microwave ion clock based on laser-cooled ${}^{113}\mathrm{Cd}^{+}$ ions. Its short-term fractional frequency stability is $4.2 \times 10^{-13}/\sqrt{\tau} $, which is close to the limit imposed by the Dick effect. After careful evaluation of various systematic frequency shifts, the 0-0 ground-state hyperfine splitting frequency of ${}^{113}\mathrm{Cd}^{+}$ was determined to be  15199862855.02799(27) Hz. This result is consist with previous measurements, and the precision has been improved to $1.8 \times 10^{-14} $, which is comparable to the performance of the mercury-ion microwave ion clock \cite{berkeland1998laser}. The uncertainty in this measurement is limited primarily by that in the frequency of the time scale with which our standard is compared. With both a better local oscillator and higher SNR transition signals, we expect in the future the uncertainty to approach $1 \times 10^{-15} $.

\begin{backmatter}
\bmsection{Funding} National Key R\&D Program of China (2016YFA0302101, 2016YFA0302103); Beijing Natural Science Foundation (1202011); Tsinghua University Initiative Scientific Research Program; National Natural Science Foundation of China (12073015).

\bmsection{Acknowledgments} We thank Kai Miao, Chenfei Wu, Liming Guo, Huaxing Hu, Wenxin Shi and Tiangang Zhao for their fruitful and helpful discussions.

\bmsection{Disclosures} The authors declare no conflicts of interest.

\end{backmatter}

\bibliography{sample}

\bibliographyfullrefs{sample}

\end{document}